\documentclass[aps,print,onecolumn,superscriptaddress,tightenlines]{revtex4}
\usepackage{amsfonts}
\usepackage{amsmath}
\usepackage{amssymb}
\usepackage{graphicx}

\begin{document}

\title{Elliptic flow of identified particles in Pb-Pb collisions at $\sqrt{%
s_{NN}}=5.02$ TeV}
\author{Er-Qin Wang}
\email{wangeq@tynu.edu.cn}
\affiliation{Department of Physics, Taiyuan Normal University, Jinzhong, Shanxi 030619,
China}
\author{Yin-Qun Ma}
\affiliation{Department of Physics, Taiyuan Normal University, Jinzhong, Shanxi 030619,
China}
\author{Li-Na Gao}
\affiliation{Department of Physics, Taiyuan Normal University, Jinzhong, Shanxi 030619,
China}
\author{San-Hong Fan}
\email{fsh729@sxu.edu.cn}
\affiliation{School of Life Science, Shanxi University, Taiyuan, Shanxi 030006, China}

\begin{abstract}
In this paper, by using a Tsallis-Pareto-type function and the
multisource thermal model, the elliptic flow coefficients of
particles $\pi ^{\pm }$, $K^{\pm }$, $p+\overline{p}$, $\Lambda +\overline{%
\Lambda }$, and $K_{S}^{0}$ produced in Pb-Pb collisions at the
center-of-mass energy of $\sqrt{s_{NN}}=5.02$ TeV are investigated. In the
process of collisional evolution, because of geometric structure,
pressure gradient, and thermal diffusion effects, deformation and
translation occurred in the isotropic emission source, leading to
anisotropy in the azimuth distribution of the final-state particles. Based
on these dynamic factors, the dependence of elliptic flow on
transverse momentum is described as well.

PACS: 14.65.Bt, 13.85.Hd, 24.10.Pa
\end{abstract}

\maketitle

\renewcommand{\baselinestretch}{1.1}

\section{Introduction}

As collision energy has gradually increased in recent years, high-energy
physics has developed rapidly. On the one hand, the energy range of
nucleus-nucleus collisions has been broadened \cite{1,2,3,4}. On the other
hand, the kinds of final-state particles measured by detectors have become
more explicit \cite{5,6,7}. This creates better conditions for obtaining a
deep understanding of the collision mechanism. The distribution of
high-energy final-state particles is important to understand the
evolutionary mechanism of fluid dynamics, where as the flow effect of
final-state particles is meaningful for the new material form, quark-gluon
plasma (QGP) \cite{8,9,10}. The formation of QGP requires an extremely
high-temperature, high-density environment. It is a state of released quarks
and gluons that is similar to an ideal fluid. From an anisotropic azimuth
analysis of final-state particles measured at the Relativistic Heavy Ion
Collider (RHIC) \cite{11} and the Large Hadron Collider (LHC) \cite{12}, it
can be seen that the generated material unaffected by gravity is QGP under
the condition of strong coupling. The quarks and gluons in the
high-temperature, high-density state are affected by multiple factors. By
means of the pressure gradient, the heterogeneity of energy density and the
asymmetry of the geometric structure at the early stage of collisions are
converted to the anisotropy of final-state particle momentum and finally
manifest as the flow effect \cite{121,122}.

In the evolutionary process of high-energy collisions, there are two main
stages, chemical freeze-out and dynamic freeze-out. The former occurs in the
formation stage of different kinds of particles, and the decay and
generation of particles remain in dynamic balance. This is an inelastic
collision process. The second process occurs later, in the diffusion stage.
Momentum and energy are maintained in a thermal equilibrium state in an
elastic collision process. After the two stages, as the temperature drops,
the final-state particles are ejected from the action system. Various
physical properties of the final-state particles are then measured by
detectors, such as the longitudinal momentum spectrum \cite{13,14}, the
rapidity (pseudorapidity) distribution \cite{15,16}, the multiplicity
distribution \cite{17,18}, and the flow effect \cite%
{19,20,101,102,103,104,105,106,107,108}. By analysis of the final-state
distribution using various theoretical models, the dynamic evolutionary
mechanism, phase graph information, and particle attribution of quantum
chromodynamics were deduced.

In non-central nucleus-nucleus collisions, the main coefficient of the flow
effect is the second-order harmonic, which is called elliptic flow ($v_{2}$%
). The value is used to represent collective motion in the system.
Collective motion is one of the characteristics formed in collisions of QGP.
The flow effect that is caused by the asymmetry of the initial geometric
structure and the heterogeneous energy of the action system includes direct
flow, elliptic flow, and triangular flow. All the harmonics are quantified
by the coefficient ($v_{n}$) of Fourier decomposition \cite{21,22}:

\begin{equation}
\frac{dN}{d\varphi }\varpropto 1+2\sum v_{n}\cos \left[ n(\varphi -\Psi _{n})%
\right] ,  \label{1}
\end{equation}

Similar long-range ridge structures and positive coefficients $v_{2}$ have
been observed in experiments \cite{19}. In theory, it is assumed that these
are based on the collective effect caused by hydrodynamic evolution of
colliding particles.

Previous studies \cite{231,232,233} have presented a description of elliptic
flow over a smaller range. Moreover, the isotropic hypothesis on the
transverse plane and the translation and expansion effects of the emission
source are used. In this paper, based on the multisource thermal model,
using the distribution of the Tsallis-Pareto-type function, and at the
center-of-mass energy of $\sqrt{s_{NN}}=$5.02 TeV, the dependence of the
elliptic flow of the identified particles ($\pi ^{\pm }$, $K^{\pm }$, $p+%
\overline{p}$, $\Lambda +\overline{\Lambda }$, and $K_{S}^{0}$) in different
centrality intervals in Pb-Pb collisions on transverse momentum is described
\cite{24}. The multisource thermal model is a statistical model that is
based on the one-dimensional string model \cite{301} and the fireball model
\cite{302} and was developed from the thermalized cylinder model \cite%
{303,304}. According to the multisource thermal model, many local emission
sources are formed along the incident direction in high-energy collisions,
and the final-state particles and jets are generated by these emission
sources. In the rest frame of an emission source, the source is isotropic,
that is, the final particles produced by the emission source are assumed to
emit isotropically. Due to differences in impact parameters, centralities,
position in space, or energy density, the emission source's temperature,
excitation degree, and particle yield ratio may vary. In comparison with
previous work \cite{231,232,233} by the multisource thermal model, not only
is the range of transverse momentum larger, but also the identification of
the final-state particles is more accurate.

\section{Model and formulation}

In this paper, using the multisource thermal model \cite{25,26,27,28,29} and
a Tsallis-Pareto-type function \cite{30,31,32,33}, the elliptic flow of
identified particles in Pb-Pb collisions is analyzed. For each source in the
multisource model, the Tsallis-Pareto-type function shows excellent
reproducibility of the spectral measurement of many particles; the form is:
\begin{equation}
\frac{d^{2}N}{dydp_{T}}=\frac{dN}{dy}Kp_{T}\left[ 1+\frac{m_{T}-m_{0}}{nC}%
\right] ^{-n},  \label{2}
\end{equation}%
where \
\begin{eqnarray}
K &=&\frac{(n-1)(n-2)}{nC\left[ nC+(n-2)m_{0}\right] },  \label{3} \\
m_{T} &=&\sqrt{m_{0}^{2}+p_{T}^{2}},
\end{eqnarray}%
where $m_{0}$ is the rest mass, $y$ is the rapidity, and $N$\ is the number
of particles. According to some non-extensive thermodynamic particle models,
the free parameter $C$, which is related to the average particle energy,
represents the mean effective temperature in the interacting system, $dN/dy$
is the particle output at different rapidity intervals, and $n$ indicate the
non-extensivity of the process, which is the departure of the spectra from
the Boltzmann distribution. After integrating for rapidity, the distribution
density function of the transverse momentum is:

\begin{equation}
f(p_{T})=\frac{dN}{dp_{T}}=N_{0}Kp_{T}\left[ 1+\frac{m_{T}-m_{0}}{nC}\right]
^{-n},  \label{5}
\end{equation}%
where $N_{0}$ denotes the normalization constant, which depends on the free
parameters $n$ and $C$. Hence, it is natural that $\int\nolimits_{0}^{\infty
}f(p_{T})dp_{T}=1$.

\bigskip Related work \cite{34} has shown that the transverse momentum
distribution of the final-state particles formed in nucleus-nucleus
collisions satisfies the Tsallis-Pareto-type function. In accordance with
the Monte Carlo method, by Eq. (5), the transverse momentum $p_{T}$ can be
extracted. In this expression, $R_{0}$\ represents random numbers uniformly
distributed on [0,1], and $p_{T}$\ can be given as:

\begin{equation}
\int_{0}^{p_{T}}f(p_{T})dp_{T}<R_{0}<\int_{0}^{p_{T}+dp_{T}}f(p_{T})dp_{T}.
\label{6}
\end{equation}%
Under the assumption of an isotropic emission source, the azimuth
distribution of final-state particles is even, and the distribution function
is:

\begin{equation}
f_{\varphi }\left( \varphi \right) =\frac{1}{2\pi }.
\end{equation}%
By the Monte Carlo method, the random number of the azimuth can be obtained
as:

\begin{equation}
\varphi =2\pi R,
\end{equation}%
where $R$ represents a random number distributed on [0, 1]. Let the beam
direction be the $Oz$ axis, and let the reaction plane be the $xOz$ plane.
Therefore, the momentum components are

\begin{eqnarray}
p_{x} &=&p_{T}\cos \varphi , \\
p_{y} &=&p_{T}\sin \varphi .
\end{eqnarray}%
\

Due to the geometric structure of the participant, the pressure gradient,
and interaction with the medium, the emission source deforms and translates
in its rest frame. Hence, an anisotropic emission source is introduced in
the multisource thermal model. To quantify the deformation and translation
of the emission source, $a_{x}$ ($a_{y}$) and $b_{x}$\ ($b_{y}$) express the
deformation and translation of the emission source along the $Ox$\ ($Oy$)
axis, $a_{x}>1$\ ($<1$) represents expansion (compression), and $b_{x}>0$\ ($%
<0$) represents translation along the positive (negative) axis. Generally,
for particles with different centrality intervals and transverse momentum,
different $a_{x}$ ($a_{y}$) or $b_{x}$\ ($b_{y}$) are obtained. As a first
approximation, the empirical relationship can be expressed as:

\begin{equation}
a_{x}=1+k_{1}\exp (-\frac{p_{T}}{\lambda _{1}})+k_{2}p_{T},
\end{equation}%
where $k_{1}$, $\lambda _{1}$, $k_{2\text{ }}$are free parameters. For
simplicity, the default is $a_{y}=1$ and $b_{x,y}=0$. Because of
deformation, the above $p_{x}$ is revised to become:%
\begin{equation}
p_{x}^{\prime }=a_{x}p_{x}+b_{x}.
\end{equation}%
Then the converted transverse momentum is:
\begin{equation}
p_{T}^{\prime }=\sqrt{p_{x}^{\prime 2}+p_{y}^{2}}.
\end{equation}%
Finally, the elliptic flow of final-state particles can be represented as:
\begin{equation}
v_{2}=\left\langle \frac{p_{x}^{\prime 2}-p_{y}^{2}}{p_{x}^{\prime
2}+p_{y}^{2}}\right\rangle .  \label{17}
\end{equation}%
\newline
{}

\section{Comparisons with experimental data}

Using the multisource thermal model, the anisotropic spectrum data of
various particles generated in Pb-Pb collisions at $\sqrt{s_{NN}}=5.02$ TeV
\cite{24}\ are studied and analyzed. The particles $\pi ^{\pm }$, $K^{\pm }$%
, $p+\overline{p}$, $\Lambda +\overline{\Lambda }$, and $K_{S}^{0}$ are
located in different centrality intervals within 0--70\% and depend on $v_{2}
$ of the transverse momentum $p_{T}$. The rapidity is in the range $%
\left\vert y\right\vert <0.5$. For particles $\pi ^{\pm }$, $K^{\pm }$, and $%
p+\overline{p}$, the measurements in hypercenter collisions (0--1\%) are
also shown.

Figure 1 shows the elliptic flow $v_{2}(p_{T})$ of meson $\pi ^{\pm }$
generated in a Pb-Pb collision at energy $\sqrt{s_{NN}}=5.02$ TeV in
different centrality intervals. The data measured by the ALICE Collaboration
in different centrality intervals are represented by different solid
symbols, and the statistical and systematic errors are both considered in
the error bar \cite{24}. The curves are fitted to results generated by the
Tsallis-Pareto-type function in the framework of the multisource thermal
model. Table 1 shows the fitted free parameters ($C$, $n$, $k_{1}$, $\lambda
_{1}$, and $k_{2}$), $\chi ^{2}$ and the degrees of freedom (dof). Clearly
the model results are consistent with the experimental data. In the
calculation, the data fitting indicates that the effective temperature $C$
increases as the centrality percentage decreases, but that the value of $n$\
remains unchanged and is assumed to be 9. It is obvious that $v_{2}$
increases with $p_{T}$\ in the low $p_{T}$\ region, and then decreases
slowly in the high $p_{T}$\ region. The transverse momentum corresponding to
the maximum value increases with increasing particle mass. This trend is
reflected in the values of $k_{1}$, $\lambda _{1}$,\thinspace\ and $k_{2}$.
Moreover, it is not hard to find that the parameter $k_{1}$\ first increases
rapidly with the centrality percentage and then slowly decreases. Finally,
the values of $\chi ^{2}/dof$\ are in a reasonable range, which is not only
affected by experimental errors, but is also related to the inaccuracy of
the theoretical calculation results.

Figure 2 shows that $v_{2}(p_{T})$ of $K^{\pm }$ in the given centrality
interval. Similarly to Fig. 1, the solid symbols also represent the
experimental data recorded by the ALICE Collaboration, and the error bar
includes the statistical and systematic errors. The curves are the results
of fitting using the Tsallis-Pareto-type function. The fitting parameters, $%
\chi ^{2}$\ and dof are also listed in Table 1. It is apparent that the
experimental data are well fitted by the model results. In the calculation,
the values of effective temperature $C$\ decrease from the central to
peripheral collisions and are systematically larger than those for particles
$\pi ^{\pm }$. As the centrality percentage increases, the values of $k_{1}$%
\ first increase rapidly, then slowly decrease, as shown in Fig. 1.

Figure 3 shows the $v_{2}$ of $p+\overline{p}$, which depends on the
transverse momentum. Figures 4 and 5 show the relationship between the
elliptic flow and the transverse momentum spectrum of $\Lambda +\overline{%
\Lambda }$ and $K_{S}^{0}$ respectively. The solid symbols are the data
points, and the curves show the model results. The fitted parameter values,
dof and $\chi ^{2}$, are included in Table 1. It is evident that the fits
are in good agreement with the experimental data. However, as shown in Fig.
4, in the given centrality interval of 60--70\%, there is a datum point
located at $p_{T}=9$ Gev/c that deviates seriously from the fitted value.
The physical mechanism underlying this deviation is not yet understood.
Similarly, when moving from central to peripheral collisions, $C$ increases,
and $k_{1}$\ increases rapidly, then decreases slowly. Overall, the model
fits the spectrum $v_{2}(p_{T})$ of identified particles measured in
different centrality intervals by ALICE in Pb+Pb collisions at approximately
$\sqrt{s_{NN}}=5.02$ TeV.

Based on the fitted results shown in Figs. 1--5, Figure 6 shows the
dependency relationship between the expansion factor $a_{x}$\ and the
transverse momentum $p_{T}$ in the given centrality interval for different
particles $\pi ^{\pm }$, $K^{\pm }$, $p+\overline{p}$, $\Lambda +\overline{%
\Lambda }$, and $K_{S}^{0}$. For a certain particle, $a_{x}(p_{T})$ are
different in different centrality intervals. The curves with maximum and
minimum dependency relationship were chosen based on Eq.(11) and are
represented by solid and dashed lines respectively. The variation trends are
similar, but the ranges are slightly different. Furthermore, as the particle
mass increases, the range also increases. Figure 7 shows the fitting
parameter $C$, which depends on the variation of centrality. When moving
from central to peripheral collisions, the effective temperature $C$\
gradually declines.

\section{Discussion and conclusions}

According to the fitted results from the above comparisons, the fitted free
parameter $C$\ is actually not the real temperature (the kinetic freeze-out
temperature) of the emission source, but the effective temperature. As is
well known, the interacting system at kinetic freeze-out (the last stage of
collision) is influenced not only by thermal motion, but also by the flow
effect. The real temperature of the emission source should reflect the
thermal motion of the particles, and therefore the real temperature of the
source is the kinetic freeze-out temperature. The effective temperature
extracted from the elliptic flow spectrum includes thermal motion and the
flow effect of the particles. By dissecting the effective temperature, it is
possible to obtain the real temperature of the interacting system. The
relationships between effective temperature, real temperature, and flow
velocity are not totally clear. Therefore, the value of effective
temperature obtained in this work is higher than the kinetic freeze-out
temperature.

Table 1 shows that the parameter $k_{1}$\ first increases rapidly with
centrality percentage and then decreases slowly. It reaches a maximum as the
centrality percentage reaches about 30\%. In addition, Fig. 6 shows that $%
a_{x}$ decreases with increasing transverse momentum $p_{T}$. However, Fig.
7 shows that the parameter $C$ declines gradually from central to peripheral
collisions. As for the dependency relationship, it can be readily understood.

From the participant-spectator geometric structure, it can be seen that as
centrality percentage increases, the extent of the overlapping parts
decreases, whereas the asymmetry rises. There is an approximate linear
relationship between the elliptic flow and the eccentricity ratio of the
participant. Hence, with increasing centrality percentage, the elliptic flow
also grows. However, $v_{2}$\ of particles in peripheral collisions is
slightly smaller than in central collisions. This may be due to shorter
system life under peripheral collisions, resulting in small $v_{2}$. Hence, $%
k_{1}$\ first increases rapidly with the centrality percentage and then
decreases slowly.

However, as the centrality percentage rises, the effective temperature $C$
declines gradually. In accordance with the geometric structure of
collisions, as the centrality percentage decreases, the number of involved
nucleons increases, and the overlapping parts also increase, leading to
higher energy density and strength of interaction, which manifests as higher
temperature. The effective temperature $C$ obtained in this study was higher
than the true temperature. The reason for this was that the effective
temperature incorporates the true temperature and the flow effect. The value
excluding the flow effect should be equal to the true temperature. Fig. 7
shows that for particles with considerable mass, the low variation ranges of
effective temperature are similar.

In short, based on the multisource model, by introducing a
Tsallis-Pareto-type function, the elliptic flow of identified particles
generated in Pb-Pb collisions at $\sqrt{s_{NN}}=5.02$ TeV was correctly
analyzed. Therefore, in the collision process, the asymmetry, expansion, and
translation effects of geometric structure affect the dynamics of the
final-state particles.

\textbf{Data Availability}

The data used to support the findings of this study are included within the
article.

\textbf{Ethical Approval}

The authors declare that they are in compliance with ethical standards
regarding the content of this paper.

\textbf{Conflicts of Interest}

The authors declare that they have no conflicts of interest regarding the
publication of this paper.

\begin{acknowledgments}
This work was supported by the National Natural Science Foundation of China
Grant Nos. 11447137 and 11575103 and the Doctoral Scientific Research
Foundation of Taiyuan Normal University under Grant No. I170108.
\end{acknowledgments}

\newpage

{\small {Table 1. Values of $C$, }}$n${\small {, $k_{1}$,}}$\lambda ${\small
{$_{1}$,$k_{2}$, $\chi ^{2}$ number of degrees of freedom (dof)
corresponding to the fits in Figs. 1--5. \\[5mm]
}}

\begin{center}
\begin{tabular}{ccccccccc}
\hline
Figure & particles & centrality & $C(GeV)$ & $n$ & $k_{1}$ & $\lambda _{1}$
& $k_{2}$ & $\chi ^{2}/dof$ \\ \hline
Fig.1 & $\pi ^{\pm }$ & 0--1\% & 1.00 & 9 & 0.17 & 2.35 & 0.001 & 4/17 \\
Fig.1 & $\pi ^{\pm }$ & 0--5\% & 1.10 & 9 & 0.27 & 2.35 & 0.001 & 6/17 \\
Fig.1 & $\pi ^{\pm }$ & 5--10\% & 1.10 & 9 & 0.49 & 2.35 & 0.004 & 2/17 \\
Fig.1 & $\pi ^{\pm }$ & 10--20\% & 0.80 & 9 & 0.60 & 2.35 & 0.004 & 3/17 \\
Fig.1 & $\pi ^{\pm }$ & 20--30\% & 0.60 & 9 & 0.65 & 2.35 & 0.004 & 2/17 \\
Fig.1 & $\pi ^{\pm }$ & 30--40\% & 0.50 & 9 & 0.64 & 2.35 & 0.004 & 2/17 \\
Fig.1 & $\pi ^{\pm }$ & 40--50\% & 0.40 & 9 & 0.59 & 2.35 & 0.004 & 7/17 \\
Fig.1 & $\pi ^{\pm }$ & 50--60\% & 0.40 & 9 & 0.54 & 2.35 & 0.006 & 1/17 \\
Fig.1 & $\pi ^{\pm }$ & 60--70\% & 0.40 & 9 & 0.48 & 2.40 & 0.005 & 11/17 \\
\hline
Fig.2 & $K^{\pm }$ & 0--1\% & 2.60 & 9 & 0.28 & 2.35 & 0.000 & 5/12 \\
Fig.2 & $K^{\pm }$ & 0--5\% & 2.20 & 9 & 0.40 & 2.35 & 0.000 & 12/12 \\
Fig.2 & $K^{\pm }$ & 5--10\% & 1.70 & 9 & 0.66 & 2.35 & 0.002 & 8/12 \\
Fig.2 & $K^{\pm }$ & 10--20\% & 1.25 & 9 & 0.86 & 2.25 & 0.002 & 4/12 \\
Fig.2 & $K^{\pm }$ & 20--30\% & 1.00 & 9 & 0.98 & 2.15 & 0.003 & 2/12 \\
Fig.2 & $K^{\pm }$ & 30--40\% & 0.72 & 9 & 0.83 & 2.35 & 0.002 & 6/12 \\
Fig.2 & $K^{\pm }$ & 40--50\% & 0.68 & 9 & 0.84 & 2.20 & 0.002 & 2/12 \\
Fig.2 & $K^{\pm }$ & 50--60\% & 0.55 & 9 & 0.64 & 2.35 & 0.003 & 1/12 \\
Fig.2 & $K^{\pm }$ & 60--70\% & 0.40 & 9 & 0.44 & 2.40 & 0.005 & 1/12 \\
\hline
Fig.3 & $p+\bar{p}$ & 0--1\% & 3.50 & 9 & 0.40 & 2.40 & 0.002 & 10/15 \\
Fig.3 & $p+\bar{p}$ & 0--5\% & 4.40 & 9 & 0.70 & 2.40 & 0.002 & 33/15 \\
Fig.3 & $p+\bar{p}$ & 5--10\% & 2.80 & 9 & 1.05 & 2.35 & 0.002 & 25/15 \\
Fig.3 & $p+\bar{p}$ & 10--20\% & 1.70 & 9 & 1.25 & 2.35 & 0.006 & 18/15 \\
Fig.3 & $p+\bar{p}$ & 20--30\% & 1.30 & 9 & 1.25 & 2.35 & 0.007 & 23/15 \\
Fig.3 & $p+\bar{p}$ & 30--40\% & 1.10 & 9 & 1.25 & 2.35 & 0.007 & 12/15 \\
Fig.3 & $p+\bar{p}$ & 40--50\% & 0.95 & 9 & 1.10 & 2.35 & 0.006 & 8/15 \\
Fig.3 & $p+\bar{p}$ & 50--60\% & 0.75 & 9 & 0.97 & 2.35 & 0.006 & 2/15 \\
Fig.3 & $p+\bar{p}$ & 60--70\% & 0.75 & 9 & 0.77 & 2.35 & 0.006 & 1/15 \\
\hline\hline
\end{tabular}

\bigskip

\bigskip

\bigskip

\bigskip

\bigskip

\bigskip

\begin{tabular}{ccccccccc}
\hline
Figure & particles & centrality & $C(GeV)$ & $n$ & $k_{1}$ & $\lambda _{1}$
& $k_{2}$ & $\chi ^{2}/dof$ \\ \hline
Fig.4 & $\Lambda +\bar{\Lambda}$ & 0--5\% & 4.20 & 9 & 0.58 & 3.00 & 0.005 &
12/7 \\
Fig.4 & $\Lambda +\bar{\Lambda}$ & 5--10\% & 3.00 & 9 & 1.20 & 2.55 & 0.007
& 3/7 \\
Fig.4 & $\Lambda +\bar{\Lambda}$ & 10--20\% & 2.10 & 9 & 1.57 & 2.30 & 0.009
& 2/7 \\
Fig.4 & $\Lambda +\bar{\Lambda}$ & 20--30\% & 1.40 & 9 & 1.60 & 2.30 & 0.009
& 1/7 \\
Fig.4 & $\Lambda +\bar{\Lambda}$ & 30--40\% & 1.10 & 9 & 1.42 & 2.40 & 0.009
& 1/7 \\
Fig.4 & $\Lambda +\bar{\Lambda}$ & 40--50\% & 0.90 & 9 & 1.26 & 2.55 & 0.005
& 1/7 \\
Fig.4 & $\Lambda +\bar{\Lambda}$ & 50--60\% & 0.80 & 9 & 1.07 & 2.50 & 0.009
& 1/7 \\
Fig.4 & $\Lambda +\bar{\Lambda}$ & 60--70\% & 0.60 & 9 & 0.70 & 2.50 & 0.005
& 4/7 \\ \hline
Fig.5 & $K_{s}^{0}$ & 0--5\% & 2.10 & 9 & 0.38 & 2.20 & 0.002 & 3/8 \\
Fig.5 & $K_{s}^{0}$ & 5--10\% & 1.70 & 9 & 0.65 & 2.20 & 0.002 & 2/8 \\
Fig.5 & $K_{s}^{0}$ & 10--20\% & 1.20 & 9 & 0.78 & 2.20 & 0.006 & 1/8 \\
Fig.5 & $K_{s}^{0}$ & 20--30\% & 0.90 & 9 & 0.83 & 2.20 & 0.005 & 1/8 \\
Fig.5 & $K_{s}^{0}$ & 30--40\% & 0.70 & 9 & 0.79 & 2.20 & 0.008 & 1/8 \\
Fig.5 & $K_{s}^{0}$ & 40--50\% & 0.60 & 9 & 0.73 & 2.20 & 0.006 & 1/8 \\
Fig.5 & $K_{s}^{0}$ & 50--60\% & 0.55 & 9 & 0.63 & 2.40 & 0.003 & 1/8 \\
Fig.5 & $K_{s}^{0}$ & 60--70\% & 0.40 & 9 & 0.44 & 2.45 & 0.005 & 1/8 \\
\hline\hline
\end{tabular}
\end{center}

%\newpage
\begin{figure}[htbp]
\centering\includegraphics[width=12cm]{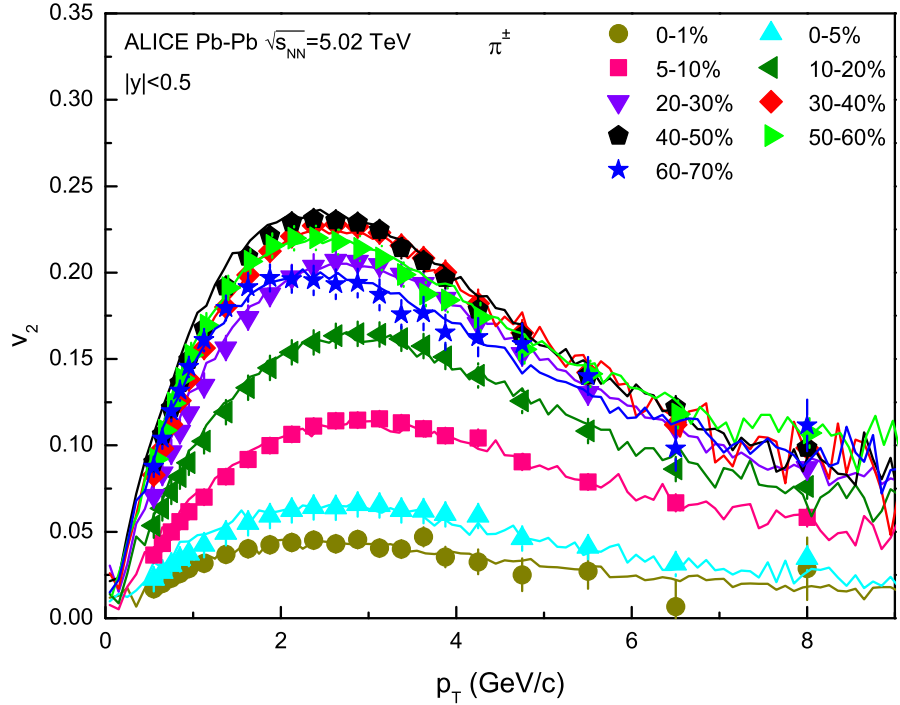}
\caption{$v_{2}(p_{T})$ of $\protect\pi ^{\pm }$ in a given centrality
interval arranged into panels of various centrality classes \protect\cite{24}%
. The data, which were measured by the ALICE Collaboration in various
centrality classes, are represented in the figure by different symbols.
Statistical and systematic uncertainties are shown as bars. The curves are
the results of this study fitted using the Tsallis-Pareto-Type function and
the multisource ideal gas model.}
\end{figure}

%\newpage
\begin{figure}[htbp]
\centering\includegraphics[width=12cm]{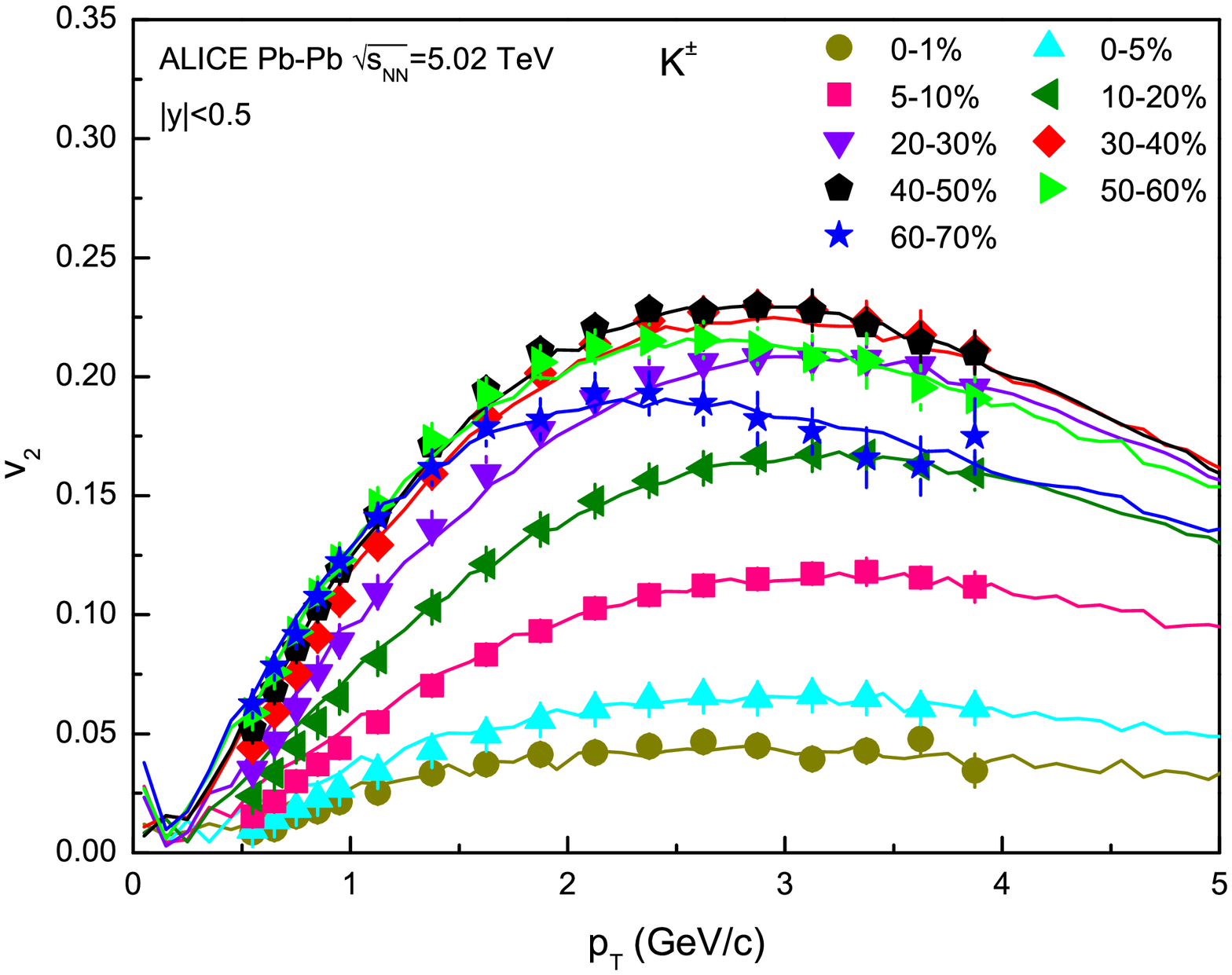}
\caption{As for Fig. 1, but showing $v_{2}(p_{T})$ of $K ^{\pm }$ for a
given centrality \protect\cite{24}.}
\end{figure}

%\newpage
\begin{figure}[htbp]
\centering\includegraphics[width=12cm]{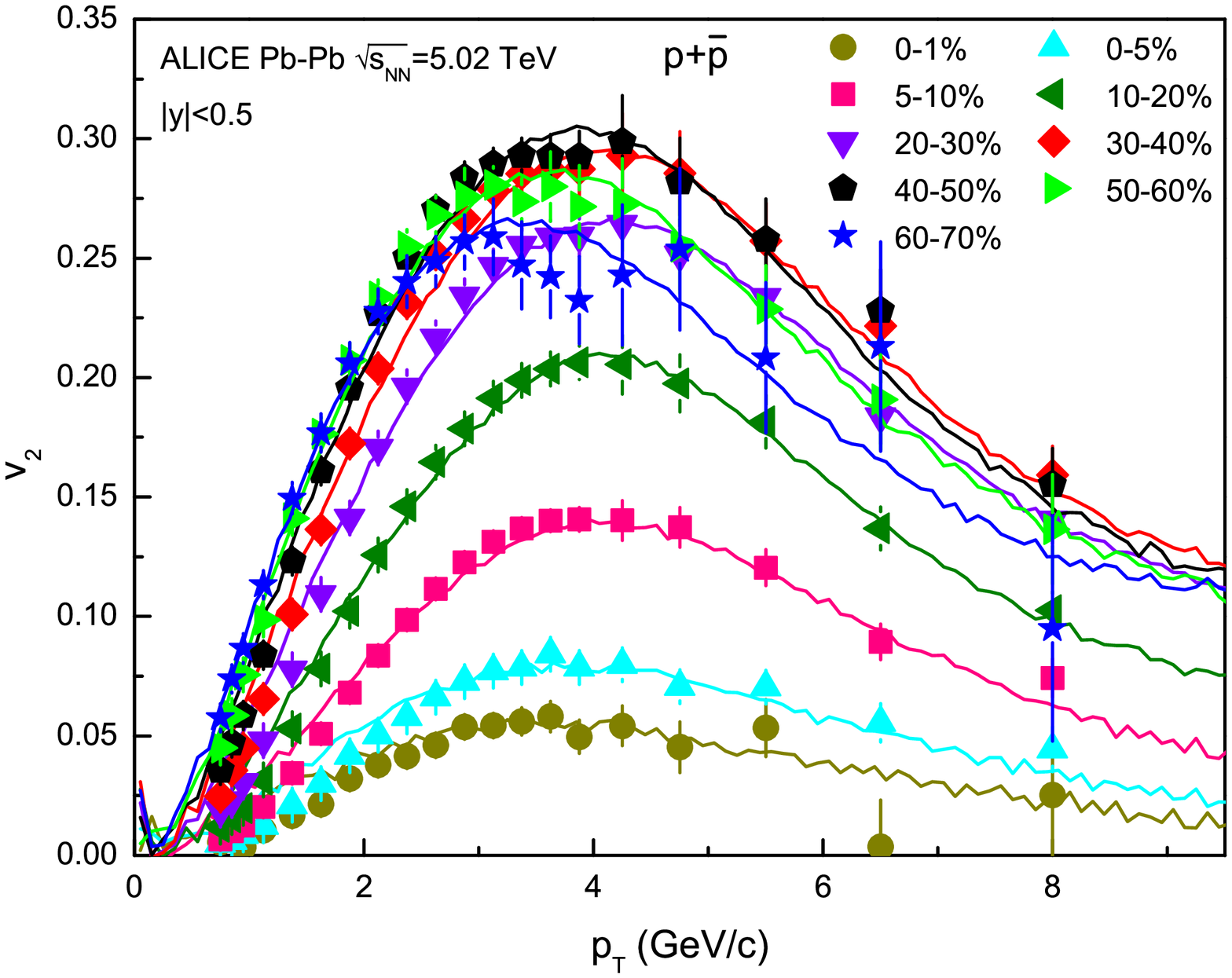}
\caption{As for Fig. 1, but showing $v_{2}(p_{T})$ of $p+\bar{p}$ for a
given centrality \protect\cite{24}.}
\end{figure}

%\newpage
\begin{figure}[htbp]
\centering\includegraphics[width=12cm]{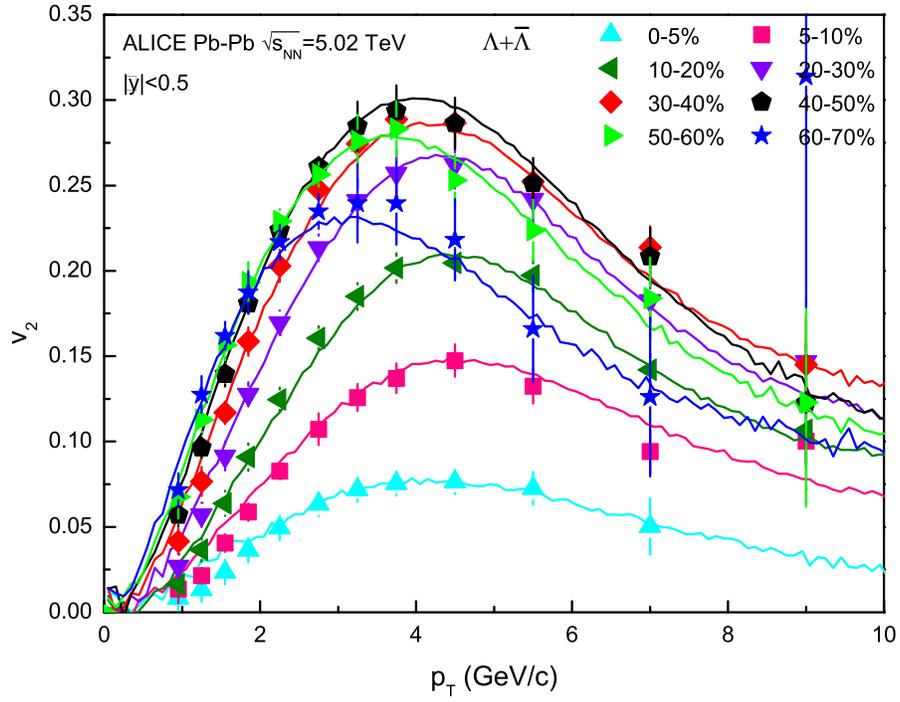}
\caption{As for Fig. 1, but showing $v_{2}(p_{T})$ of $\Lambda+\bar{\Lambda}$
for a given centrality. \protect\cite{24}}
\end{figure}

%\newpage
\begin{figure}[htbp]
\centering\includegraphics[width=12cm]{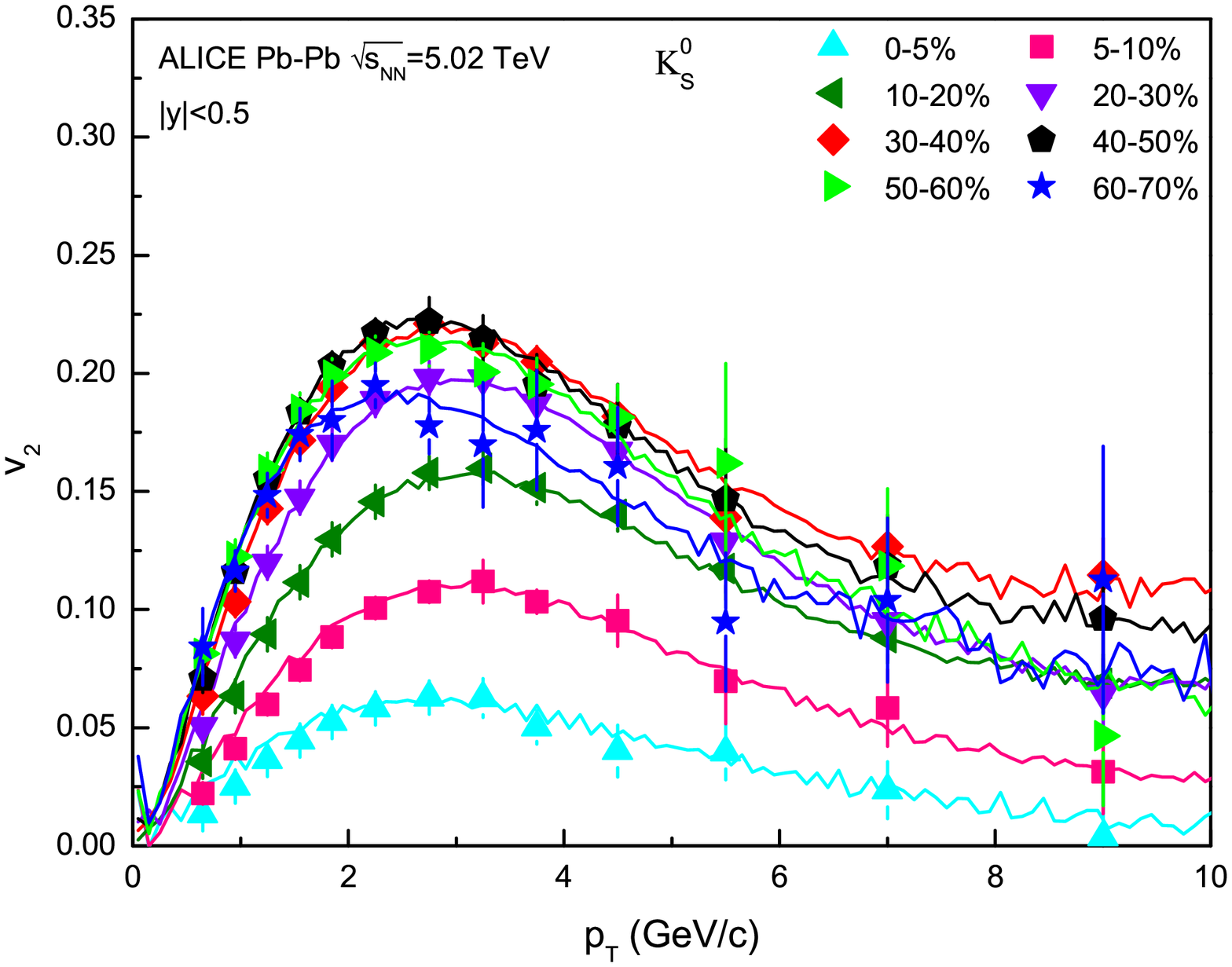}
\caption{As for Fig. 1, but showing $v_{2}(p_{T})$ of $K_{s}^{0}$ for a
given centrality. \protect\cite{24}}
\end{figure}

%\newpage
\begin{figure}[htbp]
\centering\includegraphics[width=12cm]{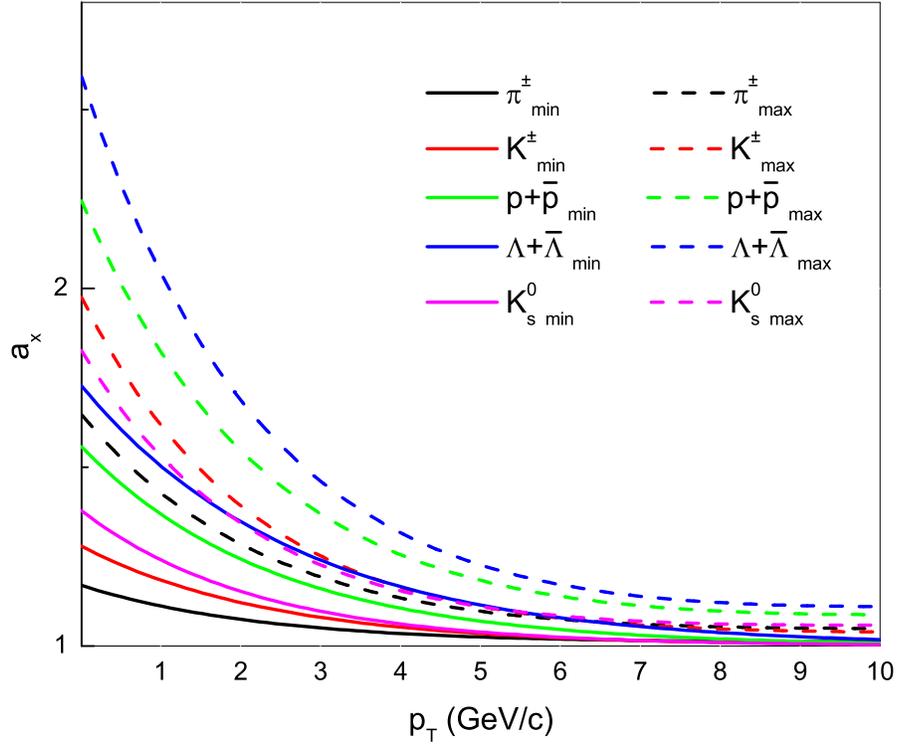}
\caption{Transverse momentum dependency on the deformation parameter $a_{x}$
of $\protect\pi^{\pm}$, $K^{\pm}$, $p+\bar{p}$, $\Lambda+\bar{\Lambda}$, and
$K_{s}^{0}$. The curves are the results of this fitted based on Eq.(11).}
\end{figure}

%\newpage
\begin{figure}[htbp]
\centering\includegraphics[width=12cm]{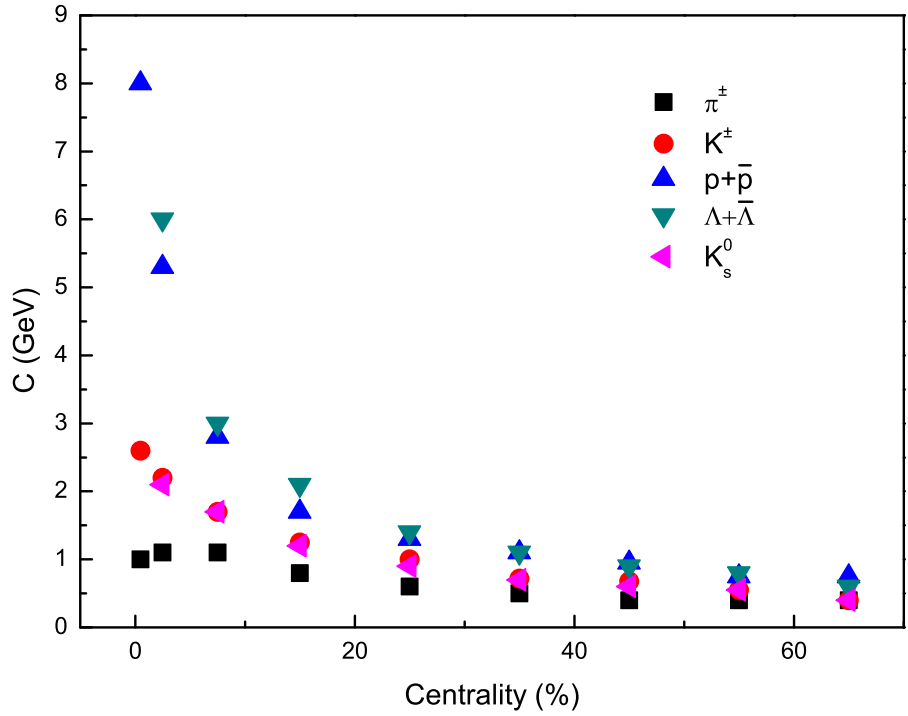}
\caption{Free parameter $C$ dependency on the centrality classes.}
\end{figure}

\end{document}